\documentclass{article}

\usepackage{PRIMEarxiv}

\usepackage[utf8]{inputenc} 
\usepackage[T1]{fontenc}    
\usepackage{hyperref}       
\usepackage{url}            
\usepackage{booktabs}       
\usepackage{amsfonts}       
\usepackage{nicefrac}       
\usepackage{microtype}      
\usepackage{lipsum}
\usepackage{fancyhdr}       
\usepackage{graphicx}       
\graphicspath{{media/}}     
\usepackage{xspace} 
\usepackage{hyphenat}
\usepackage{siunitx}
 \usepackage{amsmath}
 \usepackage{multirow}
\usepackage{subfigure}
\usepackage{appendix}
\title{On the evolution of agricultural and non-agricultural produce flow network in India
}

\author{
  Sujata S Kulkarni \\
  Discipline of Civil Engineering, \\
  Indian Institute of Technology Gandhinagar, India \\
  \And
  Raviraj Dave \\
  Discipline of Civil Engineering, \\
  Indian Institute of Technology Gandhinagar, India \\
  \And
  Udit Bhatia \\
  Discipline of Civil Engineering, \\
  Indian Institute of Technology Gandhinagar, India \\
  \texttt{bhatia.u@iitgn.ac.in}\\
  \And
  Rohini Kumar \\
  Helmholtz Centre for Environmental Research,\\
  UFZ, Leipzig, Germany\\
  \texttt{rohini.kumar@ufz.de}\\
  }

\begin{document}
\maketitle

\begin{abstract}
Rising economic instability and continuous evolution in international relations demand a self-reliant trade and commodity flow networks at regional scales to efficiently address the growing human needs of a nation. Despite its importance in securing India’s food security, the potential advantages of inland trade remain unexplored. Here we perform a comprehensive analysis of agricultural flows and contrast it with non- agricultural commodities flow across Indian states. The spatiotemporal evolution of both the networks for the period 2010 to 2018 was studied and compared using network properties along with the total traded value. Our results show an increase in annual traded volume by nearly 37\% and 87\%, respectively, for agriculture and non-agriculture trade. An increase in total trade volume without a significant increase in connectivity over the analysed time period is observed in both networks, reveals the over-reliance and increased dependency on particular export hubs. Our analysis further revealed a more homogeneous distribution between import and export connection nodes for agriculture trade compared to non-agriculture trade, where Indian states with high exports also have high imports. Overall our analysis provides a quantitative description of Indian inland trade as a complex network that could further us design resilient trade networks within the nation.
\end{abstract}


\section{Introduction}
Rapid acceleration in urbanization and intensive population growth have elevated the global demand for resources, making the trade network vital to cope with the changing market \cite{chen2018global,wang2020mapping}. The recent advances in critical infrastructure with the new trade policies have facilitated the growth of the trade network and made it more interconnected globally \cite{porkka2013food, ismail2015impact,rehman2020does}. While the international trade network provides economic leverage to participating nations, it also poses an economic and financial threat to the highly interconnected countries at times of economic shock due to external stresses, namely natural disasters and global pandemics \cite{bems2010demand,korniyenko2017assessing,min2018correlated,osberghaus2019effects,gomez2020fragility}. The few cases of such a disruption to the international trade network are the Thailand flood (2011) \cite{korniyenko2017assessing}, Japan earthquake and nuclear disaster (2011) \cite{korniyenko2017assessing, hamano2020natural}, Hurricane Harvey in the United States (2017) \cite{botzen2019economic}, and the COVID-19 pandemic (2019)\cite{verschuur2021observed}. Moreover, the sensitivity of international trade to exchange rates \cite{oguro2008trade}, transportation costs and delay \cite{mancuso2020export}, culture \cite{fidrmuc2016foreign}, and geopolitics \cite{kumar2020india} makes the network fragile, which can imperil a nation's food security and impair it's economic condition. The fragility of the international trade network highlights the importance of a localized (interstate) trade network within the country, which can reduce impacts of external shocks and minimize the damage especially in the developing nations. 
The complex network approach has garnered considerable attention from the scientific community in understanding the structural and dynamic behavior of networks in disparate domains \cite{xiao2017complex}. In case of the trade network, the complex network analysis help us in quantifying the underlying complexity of trade networks \cite{maluck2015network, wang2020mapping}. Previous work on the trade networks has mainly focused on the network at an international scale \cite{park2005recent,garlaschelli2005structure,shutters2012agricultural,lee2013applications,gao2015features,maluck2015network, xiao2017complex,wang2020mapping,herman2021modeling} or the interaction of a single commodity among and within nations \cite{gephart2015structure,du2017complex,ren2020spatiotemporal,wang2019evolution,wang2020mapping,li2021global,geng2014dynamic}. However, studies on a domestic exchange of resources in trade networks seldom exist, especially in developing countries. The significant challenge in quantifying and analyzing the domestic trade network in developing nations is the paucity of data on the trade network. Understanding the intertwined domestic trade networks helps to identify the decentralization of the supply chain system and increase efficiency in addressing the population's needs. Furthermore, analyzing the evolution of trade networks is crucial to capture the uncertainty and fluctuation associated with trading over temporal and spatial scales.

In this study, we use the complex network lens to understand the evolution of Domestic Interstate Trade Network (DITN) across India.  We select India's domestic trade network encompassing a vast array of commodities (both agricultural and non-agricultural) traded through railways for our analysis. 
India is world's 6\textsuperscript{th} largest economy with nominal Gross Domestic Product (GDP) of US\$ 2.94 trillion \cite{worldeco}. The nation is the nexus of the trade network in South Asian countries owing to it's diverse resources and an ideal geopolitical location \cite{kumar2020india}. India, being a land of diverse climatic, cultural, and socio-economic characteristics, strongly relies on interstate trade to meet the demand of raw materials for various industries and meet the requirements of residents. As a densely populated country, India needs food  and economic security that can be achieved through enhancing their internal trade by understanding the trade evolution of agriculture and non-agriculture commodities \cite{martin2017agricultural,erokhin2019handbook,xi2019impact,hu2020characteristics}. We quantify the topological features of the network to understand the structural characteristics of the trade network across two major sectors comprising commodities from agricultural and non-agricultural sectors. We examine the role of different states in the trade and their contribution to import and export in these two types of commodities over the temporal window of 2010 to 2018.  At last, to quantify the nature of domestic trade with dynamic relation of import and export in agriculture as well as the non-agriculture trade, we analyze the inward and outward movement of commodities over the spatio-temporal scale.   Our study offers new insights to understand the spatial and temporal interaction of commodities over a regional scale and hot spots of the trade network, which can further improve trade policies \cite{sajedianfard2021quantitative} and crucial for devising the resilient and recovery strategies.

We organize the rest of the manuscript as follows: In Section 2, we first discuss the details of the data sets used in the study. Then, we present the overview of the analysis, the construction of DITN, and the topological characteristics of a network. After this, we brief the evolution of the trade network over both spatial and temporal scales and compare classes of commodities. Section 4 demonstrates the results. Finally, in Section 5, we discuss the spatio-temporal behavior of interstate trade networks.

\section{Methods}
\label{section:Methods}

 \subsection{Data and Network construction}
 
We obtain the interstate movement of resources data for 2010-2018 from the Directorate General of Commercial Intelligence Statistics (DGCIS), Government of India \cite{data2021}. The DGCIS database has highly decentralized/disseminated trade data, decomposed into seventy commodities in the twenty-nine states traded through rail networks. We consider the mode of transport for trade as the railways due to the preferred mode of transporting bulk substances \cite{mukherjee2004trade}. The traded flow of resources between two states are given in quintals. The first of its kind study where we harmonized the commodities data and classified it into two main categories: agriculture and non-agriculture to understand the aggregated level trade transfer. (Table \ref{tb:Table 1}) shows the sorting of different commodities into the two categories.

 \begin{table}[ht]
\begin{center}
\caption{Classification of commodities}
\label{tb:Table 1}
\begin{tabular}{|p{3cm}||p{12cm}|} 

 \hline
 Category & Commodity \\
 \hline
 Agriculture   & rice, gram and gram products, pulses, wheat and wheat products, jowar, bajra, oilseeds, cotton, fruits, vegetables, sugarcane, fodder and husk, jute, rubber, tea, coffee, spices, livestock and products.  \\
  \hline
 Non Agriculture &   metal ores, coal, coke, cement, bricks, iron steel, soap, salt, paper, paint, varnish, chemicals, alcohols, metal products, electrical goods, caustic, potash, soda, transport equipment, machinery and tools, fertilizers, gases and other commodities.  \\
 \hline
 \end{tabular}
\end{center}

\end{table}

We aggregated the harmonized commodity-specific trade matrices to obtain the total annual agricultural and non-agricultural trade volume across the states for nine consecutive years (2010--2018). The harmonized data summarize the resource flows between states and can be represented as directed and weighted DITN. In the network, the node $i$ represents the state in India. The link $l_{ij}$ demonstrates the flow of resources between states $i$ and $j$. The direction of flow (i $\rightarrow$ j) indicates the relationship between the exporter ($i$) and importer ($j$). After construction of links, weight $w_{ij}$ (trade volume) is attributed to each link. In this study, The trade volume is considered in physical values (weight) to represent the flow between two states. Thus we construct the weighted directed trade networks each year for $N$=36 nodes, representing 28 Indian states and 8 Union territories which have pre-defined political boundaries. We analyze the DITN primarily through two aspects: the topology of the trade network and the evolutionary characteristics of interstate trade patterns.

\subsection{Topological characteristics of network}
We consider multiple metrics to characterise the topological characteristics of network. These are briefly described below.

\subsubsection{Adjacency matrix}\hfill

Adjacency matrix is concise representation of network \cite{noel2005understanding, barabasi2013network}. It is matrix of size $N \times N$ with each element $A_{ij}$=$w$ representing weight between nodes $i$ and $j$ if link exist between them, otherwise its $A_{ij}$=0. Adjacency matrix for undirected network is symmetric. 

\begin{equation}\label{eq:1}
\mathbf{A}_{i j}= \begin{cases}w, & \text { if }(i\neq j) \hspace{3em} \text {such that}  \hspace{1em}w = \{ x:x\in \mathbb{R}^{+}\} \\ 0, & \text { if } (i= j)\end{cases}
\end{equation}

The directed and weighted interstate trade network is represented by the non-symmetric adjacency matrix $A$ with each element $A_{ij}$ depicting the trade transfers between a pair of states $i$ and $j$ in quintals, where $A_{ij} \neq A_{ji}$. 

\subsubsection{Degree of Node}\hfill
 
The degree of node ($k$) is a vital parameter to express the structural property of a node in a complex network \cite{barabasi2013network}. The node degree represents the total number of links connected to a given node. Directed networks have two degrees associated with nodes, namely in-degree and out-degree, based on the links inward and outward direction, respectively. The node's degree can be calculated using adjacency matrix $A_{ij}$. 

The in-degree and out-degree of the states (nodes) in the interstate trade network are defined as (Equations~\ref{eq:2} and \ref{eq:3}):
\begin{equation}\label{eq:2}
k_i^{out} = \sum_{j=1}^n{A_{ij}}, 
\end{equation}

\begin{equation}\label{eq:3}
k_i^{in} = \sum_{j=1}^n{A_{ji}}
\end{equation}

where, $k_i^{out}$ and $k_i^{in}$ represent the out-degree and in-degree of the node $i$ (state) in interstate trade network for the number of states importing and exporting to that specific states.

The average degree of network depicts the average number of links per node. The network's average in-degree and average out-degree for directed interstate trade networks remain the same (Equation~\ref{eq:4}):

 \begin{equation}\label{eq:4}
<k_i^{in}> = \frac {1}{n}\sum_{i=1}^n{k_{i}^{in}}=<k_i^{out}>=\frac {1}{n}\sum_{i=1}^n{k_{i}^{out}}
\end{equation}

where, $<k_i^{in}>$ and $<k_i^{out}>$ denote the average in-degree and average out-degree, respectively. The $<k_i^{in}>$  and $<k_i^{out}>$ for example, represents the bulk export and import of the agriculture and non-agriculture interstate trade network. 

\subsubsection{Strength centrality}\hfill

A variation in trade volume on disparate links has a proportionate impact on the weighted trade network \cite{wang2020structure,wang2020mapping}. To understand this impact, weighted degree is calculated, which is referred to as node strength ($s_i$) \cite{barrat2004architecture}. In the weighted directed network, we use the in-strength and out-strength of a node to identify the major importer and exporters across the interstate trade network. It can be expressed as (Equation~\ref{eq:5} and 6): 

\begin{equation}\label{eq:5}
s_i^{in} = \sum_{j=1}^n{A_{ji}}{w_{ji}}
\end{equation}

\begin{equation}\label{eq:6}
s_i^{out} = \sum_{j=1}^n{A_{ij}}{w_{ij}}
\end{equation}

where, $s_i^{in}$ and $s_i^{out}$ denote the in-strength and out-strength of a state, and $w_{ij}$ is the traded volume of commodity from node $i$ to node $j$.

\subsection{Analysis of interstate trade pattern}
We use a network theory to construct indicators that help analyze the overall characteristics, distribution patterns, and closeness among the trading states in the DITN. These can be represented through following metrics. 

\subsubsection{Network density}\hfill

Network density ($D$) indicates a closer connection among the nodes of a network. In weighted and directed network $G$, with $N$ number of nodes and $L$ links, the density of network is calculated using (Equation~\ref{eq:7}) \cite{fischer1995national}. A higher network density depicts the richness of DITN.

\begin{equation}\label{eq:7}
D=\frac{L}{N(N-1)}
\end{equation}

where, $N(N-1)$ is the number of maximum possible connections in a network of size $N$.

\subsubsection{Clustering coefficient}\hfill

The clustering coefficient of DITN is indicator of tightness among the participating states in the network {\cite{watts1998collective}}. The clustering coefficient indicates the degree of connectedness between the neighbouring states connected to a same trading state. A large clustering coefficient reveals that the neighbours of the state is connected well through the trading. The clustering coefficient in directed network is calculated using (Equation~\ref{eq:8}).
\begin{equation}\label{eq:8}
C_i=\frac{L_i}{k_i(k_i-1)}
\end{equation}

where, the $C_i$ is the clustering coefficient of $i^{th}$ node with value between 0 and 1, $L_i$ is the number of links between the $k_i$ neighbours of node $i$.

The degree of clustering for whole network is captured using the average clustering coefficient $<C>$. The $<C>$ is calculated as:
\begin{equation}\label{eq:9}
<C>=\frac{1}{n}\sum_{i=1}^n{C_{i}}
\end{equation}

To analyse the evolution of DITN for both agriculture and non-agriculture commodities, we use the above mentioned topological properties of complex network at both temporal and spatial scales.




\section{Results}
\subsection{Interstate trade network}

\begin{figure}[hbt!]
\centering
\includegraphics[width=0.93\textwidth,keepaspectratio]{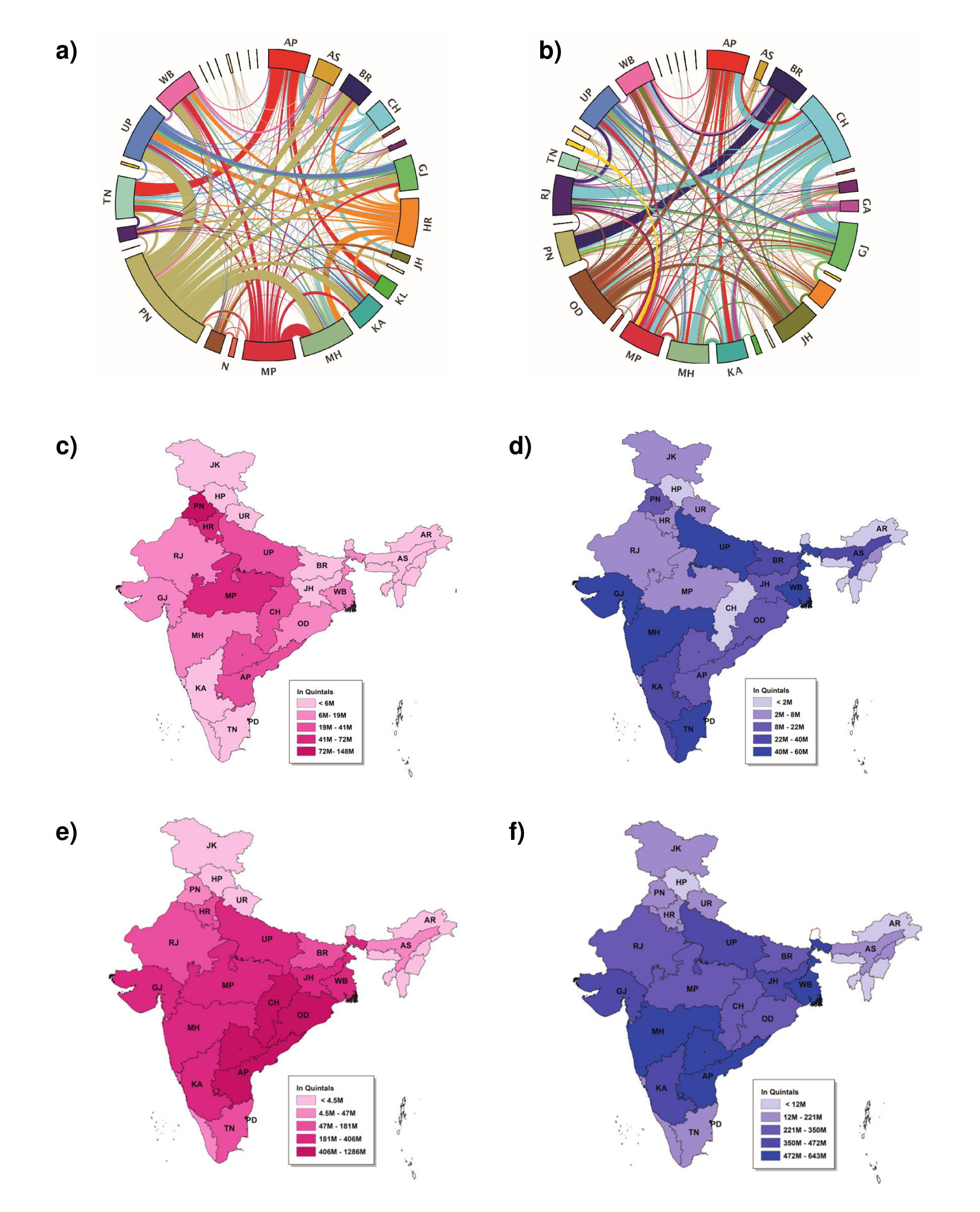}
\caption{The chord diagrams show the average trade flows between different states in India. The links' width showcases the trade volume in quintals, and the links' colors correspond to the exporting regions, (a and b) depict the trade network for Agricultural  Non-Agricultural commodities. (c and d) shows average exports and imports for Agriculture and (e) and (f) for Non-agriculture commodities in the study area. State is represented by two letters.All averages are calculated over the period of 2010-2018.}
\label{fig:Fig.1}
\end{figure}

In the DITN, we have classified the network into agriculture and non-agriculture classes based on the types of commodities. (Figure~\ref{fig:Fig.1}) shows the overview of interstate trade in India. 

\textbf {Agriculture trade:} (Figure~\ref{fig:Fig.1}a) depicts the average transfer of  agricultural and animal products across the Indian states for 2010-2018. Each year, the aggregated volume of food flow is around 1.6--2.5 million quintals. The average exports and imports of agriculture commodities of various states are presented in (Figure~\ref{fig:Fig.1}c and \ref{fig:Fig.1}d). (Table~\ref{tb:Table 2}) represents the top five top exporters and importers in the agriculture DITN along with the corresponding mean trade volumes (in quintals). The North Indian states: Punjab (PN), Haryana (HR), Uttar Pradesh (UP), and Madhya Pradesh (MP) are the key food suppliers in India, contributing with more than 70\% of agricultural products export within India. Whereas, the states like Tamil Nadu (TN), Maharashtra (MH), West Bengal (WB), Uttar Pradesh (UP), and Gujarat (GJ)  import 50\% of annual agricultural products.  

\begin{table}[ht]

\begin{center}
\caption{Top five states with highest node strengths for overall agriculture trade and non-agriculture trade}
\label{tb:Table 2}
\begin{tabular}{ |p{2cm}|p{4cm}||p{2cm}|p{4cm}| }

 \hline
 \hfil $S_{(in)}$ & weight $(quintals \times 10^6  )$ & \hfil $S_{(out)}$ & \hfil weight $(quintals \times 10^6)$\\
 \hline
 \multicolumn{4}{|c|}{Agriculture trade} \\
 \hline
 \hfil TN & \hfil 60.73 & \hfil PN & \hfil 148\\
 \hline
 \hfil MH & \hfil 56.66 & \hfil MP & \hfil 72.53\\
 \hline
 \hfil WB & \hfil 46.77 & \hfil HR & \hfil 65.85\\
 \hline
 \hfil UP & \hfil 45.15 & \hfil AP & \hfil 41.83\\
 \hline
 \hfil GJ & \hfil 41.54 & \hfil UP & \hfil 34.84\\
 \hline
 \multicolumn{4}{|c|}{Non-Agriculture trade} \\
 \hline
 \hfil WB & \hfil 643.24 & \hfil OD & \hfil 1286.14\\
 \hline
 \hfil MH & \hfil 612.04 & \hfil CH & \hfil 897.91\\
 \hline
 \hfil AP & \hfil 477.45 & \hfil AP & \hfil 701.05\\
 \hline
 \hfil KA & \hfil 468.13 & \hfil MP & \hfil 256.59\\
 \hline
 \hfil UP & \hfil 440.24 & \hfil JH & \hfil 416.76\\
 \hline

 \end{tabular}
\end{center}
\end{table}

\textbf {Non-Agriculture trade:} The non-agricultural commodities mainly include infrastructure supportive commodities like cement, mineral ores, and chemicals. The total trade volume of non-agriculture commodities in DITN is around 45-60 million quintals, being significantly larger than the agriculture products' trade. (Figure~\ref{fig:Fig.1}b) demonstrates the transfer of major non-agriculture products across the Indian states for 2010-2018. (Figures~\ref{fig:Fig.1}e and \ref{fig:Fig.1}f) show the mean exports and imports of non-agriculture goods. Chhattisgarh (CH) and Odisha (OD) contributes the 35\% exports of total non-agriculture trade in DITN, at the same time, WB, and MH are leading importers (Table \ref{tb:Table 2}).  
    
Summarizing above, overall the northern Indian states Punjab (PN) and Haryana (HR) dominate the agriculture DITN, while the south Indian states Tamil Nadu (TN) to Andhra Pradesh (AP) is the most prominent link for trade transfer with the average trade volume of 2.03~$\times$~10$^8$~quintals~yr$^{-1}$, which accounts for 4.6\% of the total trade volume of network (Figure~\ref{fig:Fig.1}a). In contrast to the agriculture trade pattern, the non-agriculture trade pattern does not exhibit a dominant state or link. The states considered for this analysis significantly participate in transferring non-agricultural commodities, albeit unevenly. The flow of non-agricultural goods from CH to PN is the link with high trade transfer the mean annual trade volume of 4.07~$\times$~10$^7$~quintals~yr$^{-1}$ comprising of 4.7\% of total non-agriculture interstate trade (Figure~\ref{fig:Fig.1}b). 

\subsection{Temporal evolution of the interstate trade network}

\begin{figure}[ht]
\centering
\includegraphics[width=1.0\textwidth,keepaspectratio]{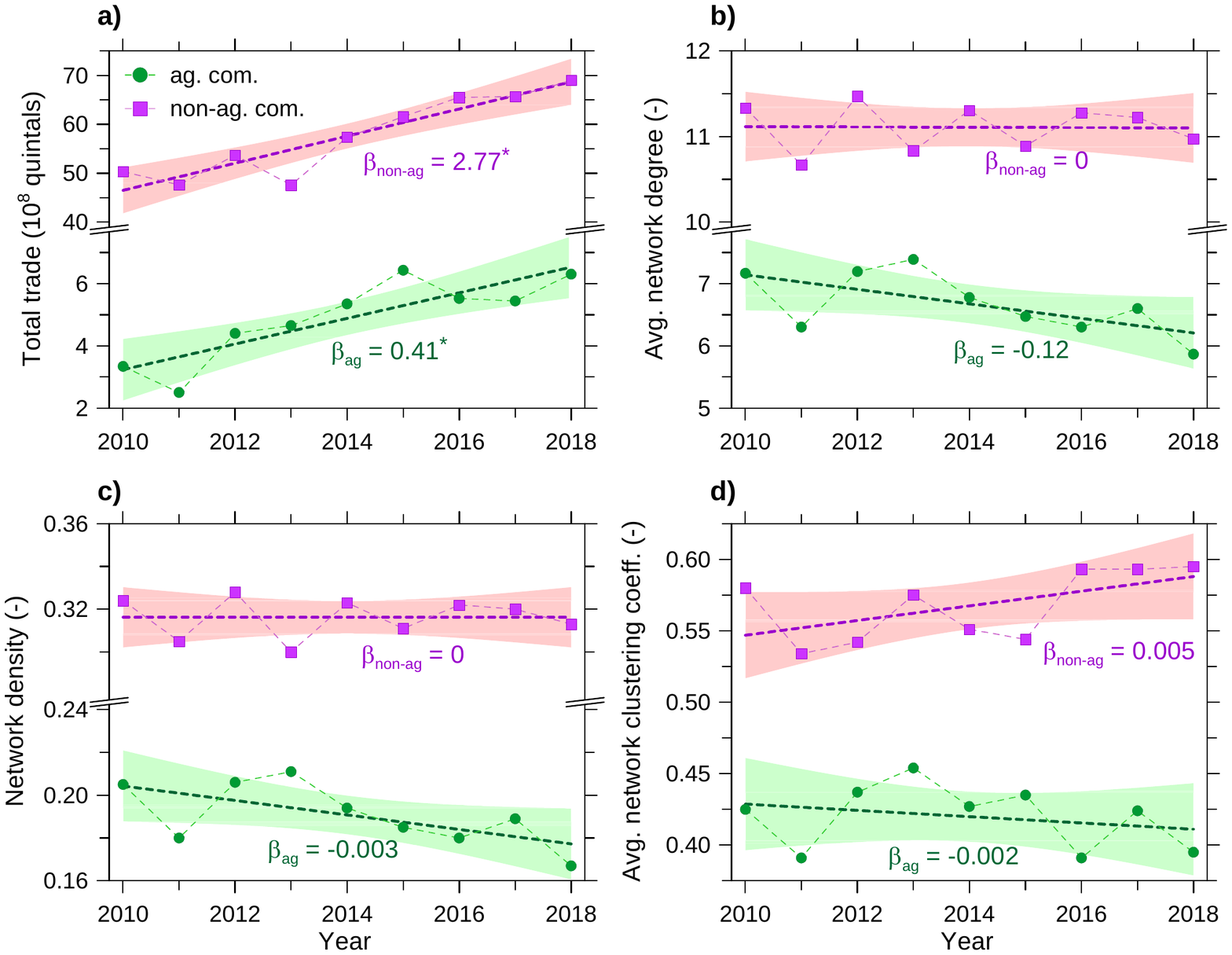}
\caption{Temporal structural evolution of total trade volume and network characteristics: (a) Total traded value (in $10^8$ quintals), (b) Average degree, (c) Network density, and (d) Average Clustering Coefficient for 2010-2018. here, $\beta$ indicates the slope of a linear trend line fitted to scattered points (values with * represent significant trend at $p-value < 0.05$). The results are presented for the Agriculture (green) and Non-Agriculture (purple) interstate trade networks.}
\label{fig:Fig.2}
\end{figure}

To understand the structural evolution of DITN on the temporal window of 2010--2018, we use the topological parameters of a complex network as detailed above (See Section ~\ref{section:Methods}). Over the last decade, the number of nodes participating in agriculture (and non-agriculture) trade increased to from 26 (29) in 2010 28 (31) in 2018, respectively, suggesting that participation in interstate trading has moderately increased. In the case of links, the non-agriculture trade has a constant number of links, whereas the links in agriculture trade have substantially increased from 223 to 264 in the time frame of 2010-2018. This signifies the increasing connectivity for agricultural trades between participating states. (Figure~\ref{fig:Fig.2}a) describes the evolution of agriculture and non-agriculture interstate trade in India. The traded volume of agriculture and non-agriculture trade increased rapidly, exhibiting the significant ($p-value < 0.05$; linear trend) evolution with time. The total traded volume increased by 86.6\% and 37.2\%, respectively, for agriculture and non-agriculture trade, respectively. The strong trend ($\beta_{non-ag}= 2.8~\times~10^8$~quintals~yr$^{-1}$) is observed for the non-agriculture trade, whereas agricultural trade has gradually increased with a dip (16.7\%) in recent years. This decrease may mark the impact of 2016-2018 droughts in India \cite{mishra2021unprecedented}. 

The topological properties of the network do not significantly change over the temporal evolution of trade. (Figure~\ref{fig:Fig.2}b) elaborates the change of average degree over time for agriculture and non-agriculture trade. The average degree of non-agriculture trade remains unchanged, insisting that the network is stable with an increase in overall trade volume. In the agriculture trade, the slight decrease ($\beta_{non-ag}= -0.12$) in the average degree with an increase in trade flow emphasizes that the network is moving towards self-reliance under the assumption that the states have not opted for the exchange of goods by other modes of transport. Note that here in our analysis, we have considered only trades via railways which is the preferred mode of transport of bulk substances -- although the neighboring states may prefer to choose other transportation modes (via roads). The network density (Figure~\ref{fig:Fig.2}c) displays the weak negative trend in agriculture trading ($\beta_{non-ag}= -0.003$). while in the case of non-agriculture trade, it remains unchanged.  Increasing total trade with no significant change in network degree depicts the diversification of commodities over the network's existing connections. Along with increasing trend in non-agricultural commodities, observing positive trend of clustering coefficient shows that the connectedness of export lines is improving over time (Figure~\ref{fig:Fig.2}d). 

\subsection{Evolution of the core states’ trade relations in the interstate trade network}
\begin{figure}[ht]
\centering
\includegraphics[width=1.0\textwidth,keepaspectratio]{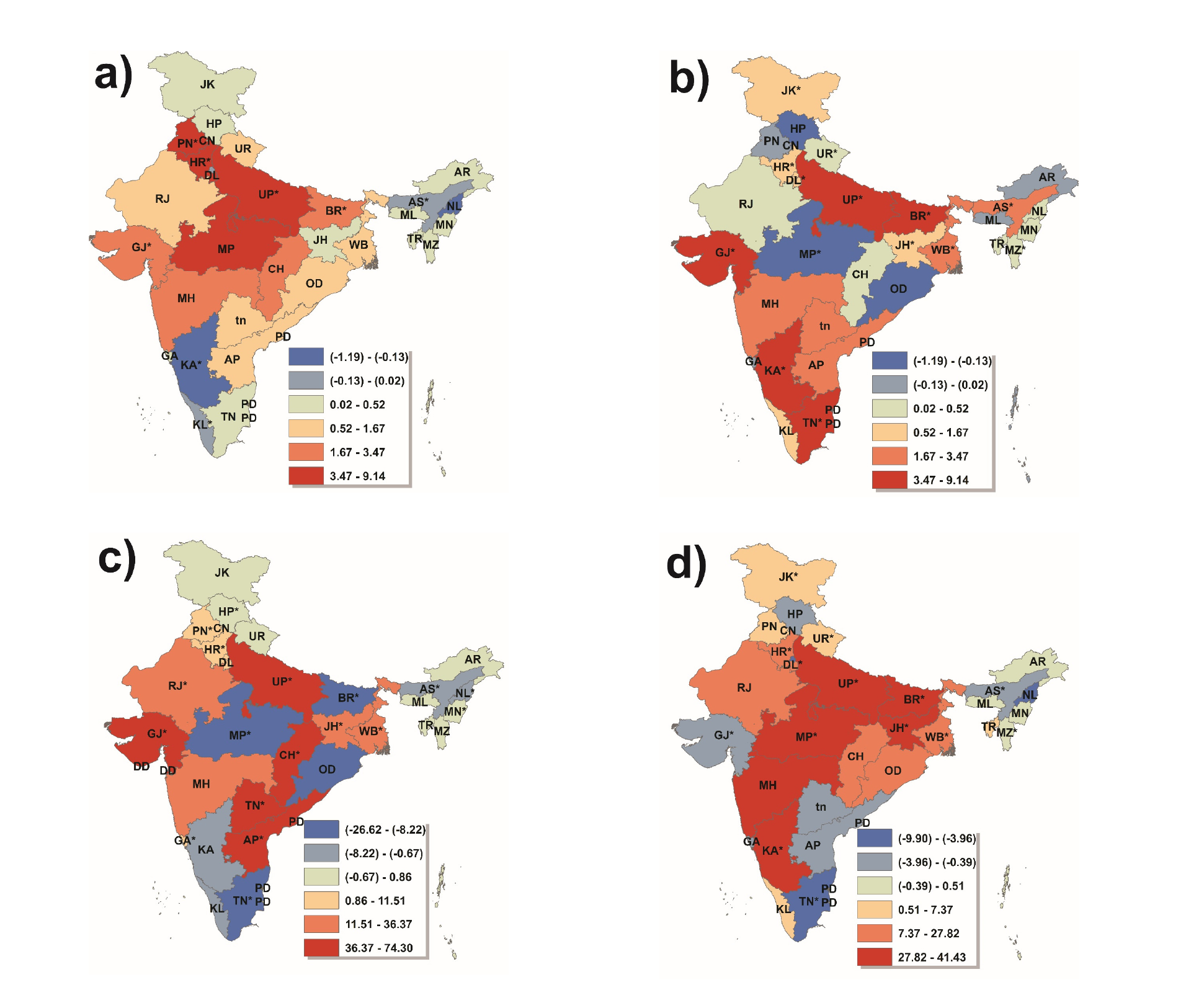}
\caption{The spatial variation in the trend of exports and imports of Agriculture (a \& b) and Non-agriculture (c \& d) commodities over the period of 2010-2018.All values are in Million Quintals.}
\label{fig:Fig.3}
\end{figure}

To identify the influence and the changes of the interstate trading for the different states, we analyse the spatial variation of the trend for import and export in agriculture and the non-agriculture trade network. (Figure~\ref{fig:Fig.3}a) emulates the spatial variation in exports of agriculture commodities. The major exporters of agricultural products such as PN, HR, UP, MP and Delhi (DL) have shown significant increase in exports. The states namely Karnataka (KA), Nagaland (NL), Kerala (KL), and Assam (AS) displayed a strong negative trend in exporting agriculture products. In the case of import of agriculture commodities, MP is the only state that has reflected negative trend with increasing export (Figure~\ref{fig:Fig.3}b). The number of states having a negative trend in the export of non-agriculture commodities is high compared to agriculture commodities (Figure~\ref{fig:Fig.3}c). On the other hand contrasting behavior is observed in the import of non-agriculture commodities as the number of states with positive trends in non-agriculture import is more than the import of agriculture products(Figure~\ref{fig:Fig.3}d). 

\begin{figure}[ht]
\centering
\includegraphics[width=1.0\textwidth,keepaspectratio]{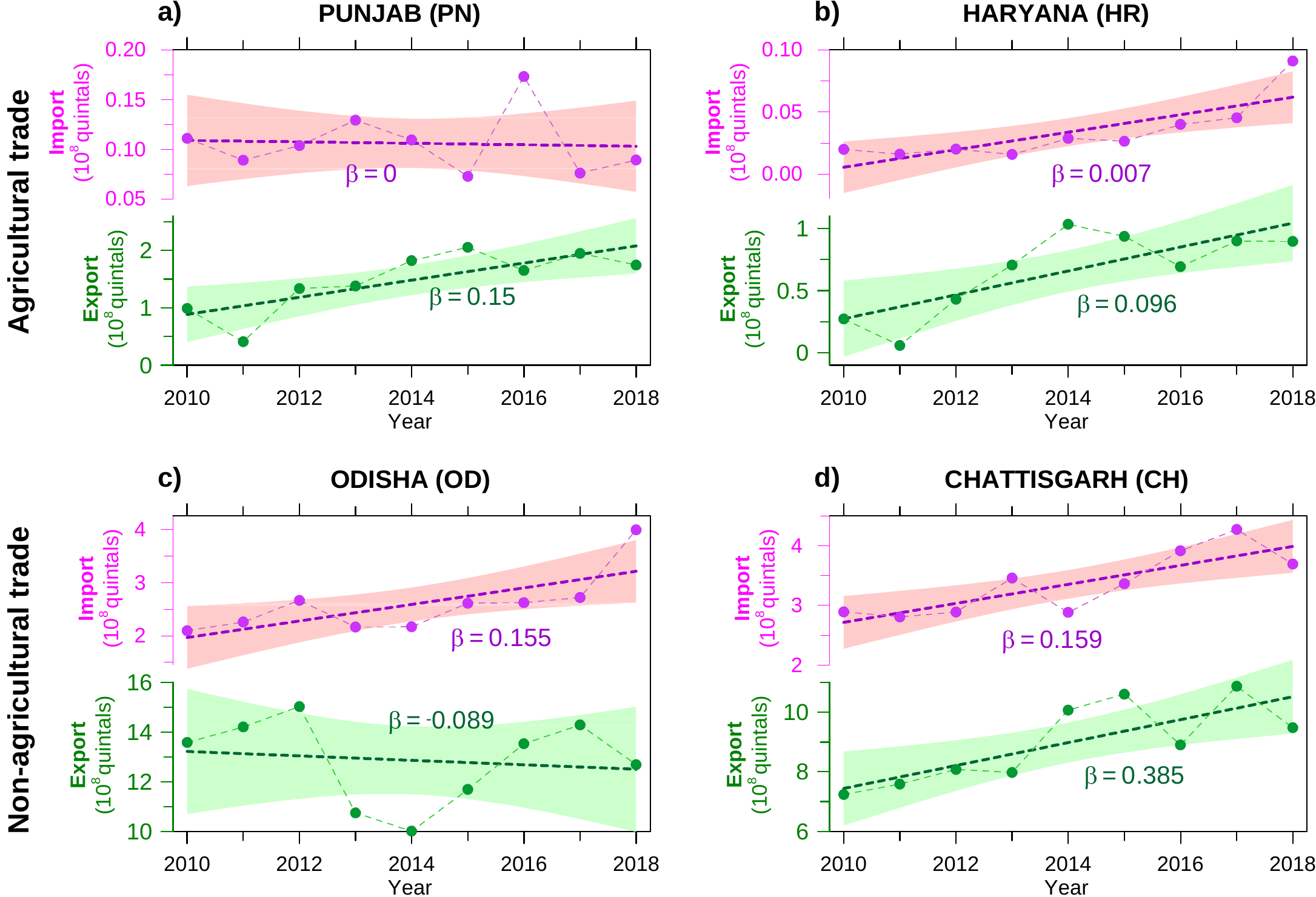}
\caption{The temporal variation of exports and imports of leading exporters of agriculture (Punjab and Haryana) and non-agriculture (Odisha and Chattisgarh) commodities over the period 2010--2018. Also shown are the respective linear regression slope ($\beta$).}
\label{fig:Fig.4}
\end{figure}

(Figure~\ref{fig:Fig.4} a and b) presents the evolution of core states that have contributed significantly to the exporting of agriculture commodities. PN and HR have steady evolution as the trends for import and export have not changed significantly over time making them more stable. While leading exporters in non-agriculture commodities, OD and CH have increase in both imports and exports.The result can be attributed to the states being pioneers in processing metallurgical-dominated products, which require the high import of raw non-agriculture products and can export the finished non-agriculture products \cite{industry2022}. (Figure~\ref{fig:Fig.4} c and d). 

\subsection{Dynamic relation of Agriculture and Non-agriculture trade in interstate trade network}
\begin{figure}[ht]
\centering
\includegraphics[width=1.0\textwidth,keepaspectratio]{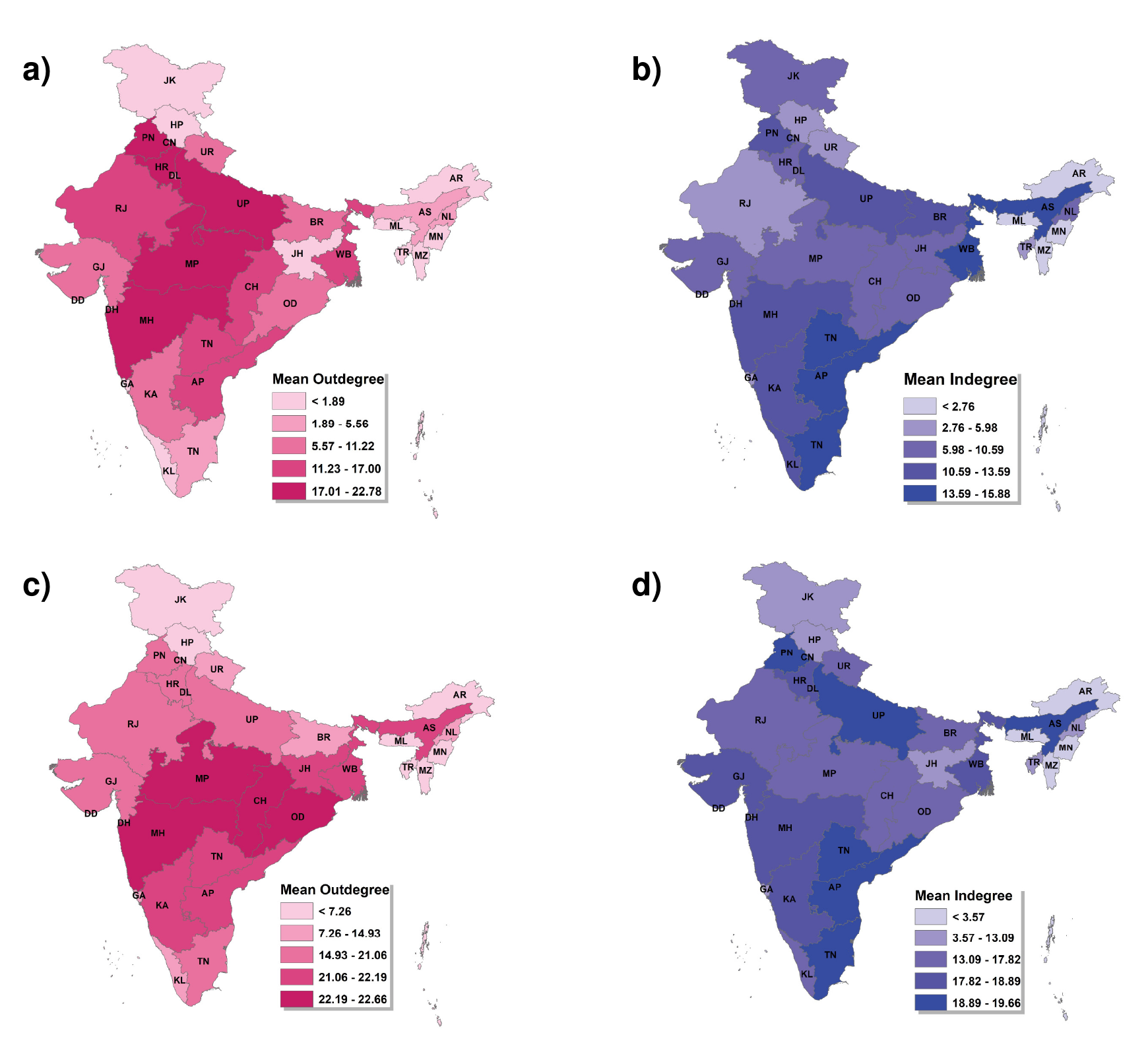}
\caption{ Spatial and temporal variations of indegree and outdegree. Study area showing average outdegree and average indegree respectively for agriculture (a \& b) and non-agriculture (c \& d) commodities.All averages are calculated over the period of 2010-2018}
\label{fig:Fig.5}
\end{figure}

Finally, to understand the evolution and dynamic relation of agriculture and non-agriculture trade with the DITN analysis, we use the in-degree and out-degree parameters' with respect of their spatial and temporal variations. (Figure~\ref{fig:Fig.5} a and b) depicts an increase in number of outgoing and incoming connections over the analysed time-frame. This is more prominent for the agriculture trade which suggests that over the period of time, the north-central belt of India has shown improvement in the export trade connection. The non-agriculture trade network, on the other side, has shown the clustering pattern in the export and import connections (Figure~\ref{fig:Fig.5} c and d) that suggests the non-agriculture trading is influenced by their states geographic location. For example, (Figure~\ref{fig:Fig.5}c) shows that the central and east central belt of India with the state (including OD,CH MP, and MH) have shown an increase connection in export where there is a cluster of the south-eastern part shown a positive trend in import connection.    

\begin{figure}[ht]
\centering
\includegraphics[width=1.0\textwidth,keepaspectratio]{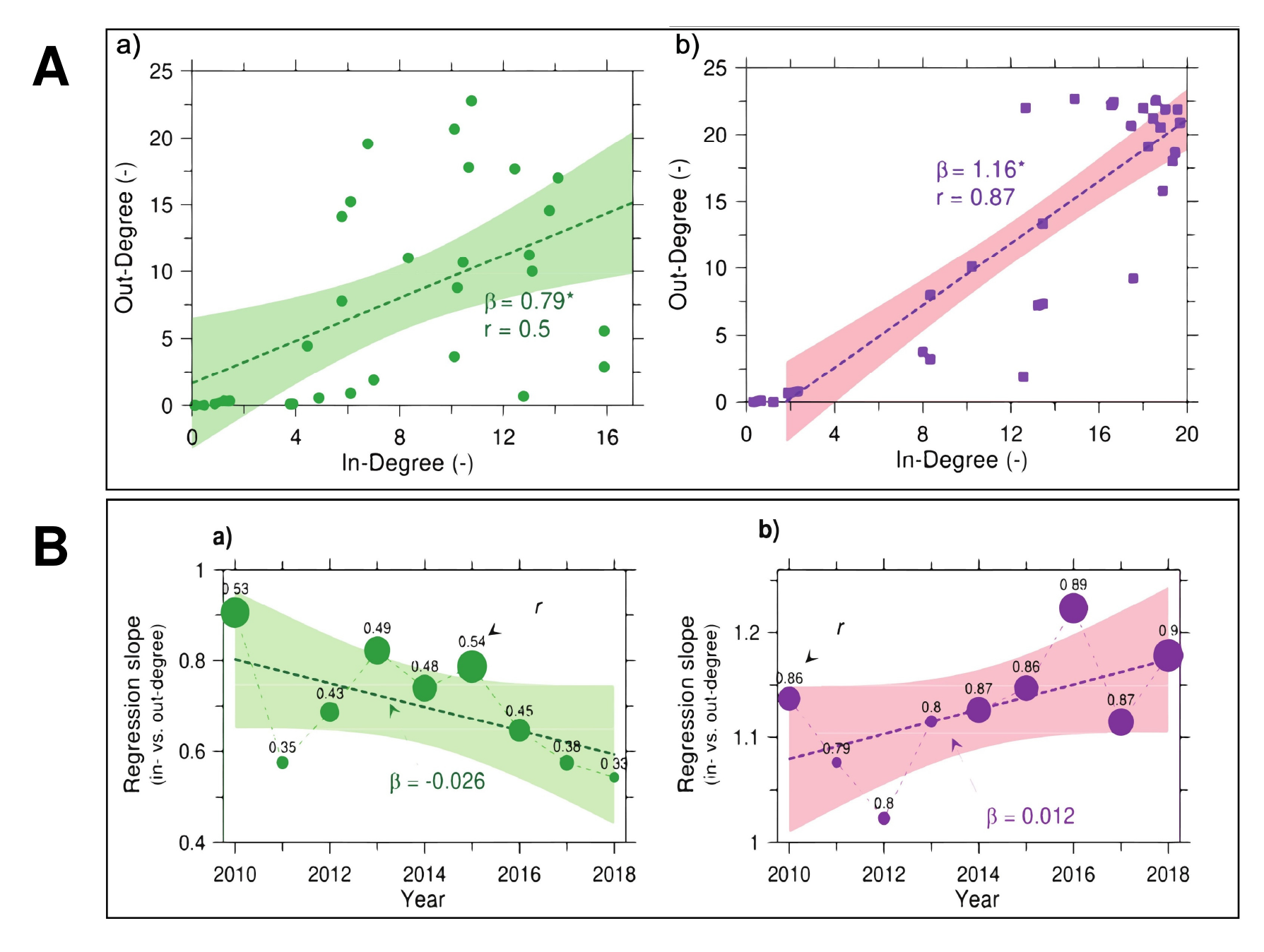}
\caption{(A) Scatter plots of average indegree and outdegree over nine years (2010-2018) for Indian states. $\beta$ indicates the slope of the fitted linear regression line to scattered points; and $r$ indicates the Pearson correlation coefficient. (B) Scatter plots of the year-wise slope of indegree and outdegree. The size and numerical above the marker represents the respective correlation value. In all panels, shown are the fitted regression line (dash line) and corresponding 95\% confidence band. The results in sub-panels (a) and (b) are for agriculture trade and non-agriculture trade, respectively.}
\label{fig:Fig.6}
\end{figure}

(Figure~\ref{fig:Fig.6}A) shows the relation of import and exports of agriculture and non-agriculture commodities over the 2010--2018 time frame based on the average in-degree and out-degree as parameters. The distribution of trade by states in agriculture product is sparse ($\beta=0.79$) and also have low correlation ($ r=0.5$) that indicates that there is no major hot-spots of states who govern the connections in agriculture trade network (Figure~\ref{fig:Fig.6} A-a). However in the non-agriculture trade, the states with high export connections also have high import connections ($\beta=1.16$ and $ r=0.87$) making them strongly connected in the respective trade networks and hot-spots for trading (Figure~\ref{fig:Fig.6} A-b).

(Figure~\ref{fig:Fig.6}B) indicates the annual evolution of network with respect to it's import and export connection, represented here as corresponding regression slope between in-degree and out-degree of network. The agriculture interstate trade network is showing negative trend ($\beta= -0.026$) in connections and also have low correlation values over the time frame. This indicates that the states are not dependent on hot-spot states and the states are becoming more self-reliant over the years through diversification, possibly to reduce the future food risk (Figure~\ref{fig:Fig.6}B - a). In contrast, the non-agriculture interstate trade network has maintained a slight positive trend in making more connections over the years. In general, with an overall high correlation value throughout the temporal window, the non-agriculture trade network appears to be highly connected and dependent on the hot-spots making them more vulnerable to disruption (Figure~\ref{fig:Fig.6}B - b).

\section{Discussion and Concluding Remarks}
The advancement in transport infrastructure and liberal trading policies have enhanced the spatio-temporal import and export at an international scale \cite{gani2017logistics,IMF2022}. The community has access to the resources they cannot produce and can procure the commodity through trade, leading to over-dependence on global trade. However, rising economic instability, disasters, and geopolitical situation infuse the fragility in the international trade network \cite{puma2015assessing, kumar2020india}. Thus, maintaining the proper domestic interstate trade network is essential to reduce impacts of such external shocks. This study evaluates the evolutionary characteristics of agriculture and non-agriculture Domestic Interstate Trade Network from 2010 to 2018 through a complex network approach. The key contributions of the study going beyond the previous studies \cite{sharma1997environmental, kumar2011export, shutters2012agricultural, maluck2015network, nag2016emerging, thomas2019competitiveness, wang2020mapping,katyaini2021water} are to explore the dynamics and nuances of DITN and the spatio-temporal evolution of the agriculture versus non-agriculture trade that provides crucial insights for authorities and policymakers to manage the interstate trading.

While International trade networks and single commodity global trade networks have received significant attention \cite{smith1992structure, gephart2015structure, del2017trends, chen2018global, wang2019evolution, pacini2021network}, the single commodity interstate trade network is rarely focused \cite{rioux2019economic, tintelnot2018trade,dhyne2021trade,chen2022china}. Understanding the evolution of interstate trade networks of multiple commodities can help us gain insights on nation's self-sufficiency and identification of hubs that are the most critical to sustain the supply chain. 
Here, we have assessed how interstate trade networks of diverse resources have evolved across India over the last decade. Our result demonstrated that non-agriculture and agriculture trade behave differently. In the case of non-agriculture DITN, no significant network density change with an increase in clustering coefficient signifies the improvement in connectedness over time, and the major trading states are relatively close in the existing trading route. In contrast, agriculture DITN  with a modest decrease in network density and clustering coefficient with increased trade depicts the trade moving towards self-reliance. The finding can be attributed to the quick residence recovery of trade due to the self-reliant agriculture trade network during external shocks and pandemic situations where the interstate movements were minimal.  

Our analysis enables us to quantify the contribution and evolution of core nodes and the connection of trade over a temporal window in agriculture and non-agriculture DITN. We note that importing non-agriculture commodities has more positive spatial growth than agriculture commodities due to economical growth in the sector and that enhanced the strength between the existing linkages of trade \cite{Nitiayog2002, budget2017india,Nitiayog2017}
. This signifies more network dependence on non-agriculture exporting states over a temporal period. Whereas in agriculture trade growth, export is limited to the northern belt of India. In importing non-agriculture commodities, the connections and evolution show a clustering pattern that depicts the geographic influence in trading due to their . The non-agriculture trade is moving towards over-dependency on the core states in trade; on the contrary, the agriculture trade is less dependent on core states. 

In order to strengthen the interpretation of our analysis, future work can address the following limitations. (a) We consider the commodities transfer through railways. Though most of the transfer of commodities are traded through railways, the consideration of commodities through air, water, and road transport will make the analysis more robust. (b) Here we have considered harmonized data of agriculture and non-agriculture commodities. However, studying the individual commodity transfer (i.e., rice, wheat, coal, and metal products) and their evolution can bring new insights. 

Trade globalization has shown an increase in food resilience and water security, whereas it also enhanced the likelihood of global crises due to interdependence. Overall, the topological quantification of Indian inland trade through a complex system's lens could help to understand the trade network's resilience and recovery through identification of more vulnerable section of network to external disruptions. The study of interstate trade at a local scale can be extended to understand virtual water trade. The virtual water traded through agriculture and non-agriculture commodities with the topological properties of the network and their evolution can help understand impacts on the local water system and aid in making informed decisions on the issue of water security \cite{konar2011water, kumar2011export,d2019global, graham2020future,nishad2022virtual}. 

\section*{Data and Code availability}
The data sets used in this study are obtained from Ministry of Commerce and Industry {\cite{data2021}}. For network analysis, NetworkX (\url{https://networkx.org}), an open-source Python package is used. 

\section*{Author contributions}
SK, RK and UB designed the experiments. SK performed the analysis. RD, RK and UB wrote the manuscript with inputs from SK. SK and RD contributed equally.

\section*{Acknowledgments}
The authors acknowledge Angana Borah, Divya Upadhyay, Pravin Bhasme and Shekhar S Goyal for active discussions and constructive comments on the manuscript. UB acknowledges IITGN Startup grant and MHRD/IISc STARS ID-367 for funding support.

\section*{Abbreviations}
DITN: Domestic Interstate Trade Network,
DGCIS: Directorate General of Commercial Intelligence Statistics,
AP: Andhra Pradesh,
AR: Arunachal Pradesh,
AS: Assam,
BR: Bihar,
CH: Chhattisgarh,
CN: Chandigarh,
DD: Daman and Diu,
DL: Dehli,
GA: Goa,
GJ: Gujarat,
HP:Himachal Pradesh,
HR: Haryana,
JH: Jharkhand,
JK: Jammu and Kashmir,
KA: Karnataka,
KL: Kerala,
MH: Maharashtra,
ML: Meghalaya,
MN: Manipur,
MP: Madhya Pradesh,
MZ: Mizoram,
NL: Nagaland,
OD: Odisha,
PN: Punjab,
RJ: Rajasthan,
SK: Sikkim,
TN: Tamilnadu,
TR: Tripura,
TS: Telangana,
UP: Uttar Pradesh,
UR: Uttarakhand,
WB: West Bengal

\clearpage
\bibliographystyle{unsrt}  
\bibliography{references} 

\newpage


\end{document}